\documentclass[10pt,notitlepage,pra,twocolumn]{revtex4}
\usepackage{amsfonts}
\usepackage{amsmath}
\usepackage{amssymb}
\usepackage{graphicx}
\usepackage[toc,page,header]{appendix}

\providecommand{\U}[1]{\protect\rule{.1in}{.1in}}

\begin{document}

\title{Quantum phases of strongly interacting Rydberg atoms in triangular
lattices}
\author{Jing Qian, Lu Zhou and Weiping Zhang}
\affiliation{Quantum Institute for Light and Atoms, Department of Physics, East China
Normal University, Shanghai 200062, People's Republic of China}

\begin{abstract}
We present a theoretical study on the system of laser-driven strongly
interacting Rydberg atoms trapped in a two-dimensional triangular lattice,
in which the dipole-dipole interactions between Rydberg states result in
exotic quantum phases. By using the mean-field theory, we investigate the
steady state solutions and analyze their dynamical stabilities. We find that
in the strong-interaction limit, the dynamics of the system is chaotic and
exhibits random oscillations under appropriate laser detunings. Lyapunov
exponent criterion is introduced to confirm the existence of this chaotic
behavior. In addition, we present a full quantum calculation based on a
six-atom model, and find that the system exhibits some bi-antiferromagnetic
properties in every triangular cell when the one-photon detuning is exactly
resonant or blue-shifted.
\end{abstract}

\maketitle
\preprint{}

\textit{Introduction}: Ultracold atomic gases loaded in an optical lattice
can provide a clean and highly controllable platform to study various
phenomena in condensed matter physics \cite{Bloch05,Bloch08}. Due to the
remarkable control of physical parameters, such as the hopping rate or the
on-site interaction strength, one can not only simulate the quantum
many-body physics of solid-state systems, but also explore exotic quantum
phases in the parameter regimes inaccessible to real solid-state materials.
Recently, two new developments have taken place in this area. Firstly,
experimental techniques for the preparation of dipolar quantum gas have been
rapidly advancing over the last few years. This has been demonstrated by the
realization of Bose-Einstein condensation of chromium atoms which have large
magnetic dipole moments \cite{Axel05} and by the creation of quantum
degenerate states of Rydberg atoms which have very large induced electric
dipole moments \cite{Rolf08,Low12}. Secondly, optical lattices with
non-cubic geometrical configurations have been introduced in this area \cite%
{Chen10,Dauphin12}. Typical examples include the experimental observation of
the superfluid to Mott insulator phase transition of rubidium atoms in
triangular optical lattices \cite{Becker10}, and quantum simulation for
triangular magnetism with the motional degrees of freedom of atoms \cite%
{Struck11}.

Due to its long-range nature, the dipolar interaction between atoms should
lead to novel kinds of quantum phases in the strongly interacting regime 
\cite{Glaetzle12}. A typical example is bistable or non-equilibrium phase
caused by dipole blockade effect in Rydberg atom gas \cite{Tony11}.
Moreover, the dipolar interaction coupled with the geometry of the
triangular lattice can produce more interesting effect named geometrical
frustration \cite{Moessner06,Eckardt10,Kim10}. A rich variety of possible
quantum phases arise from it, such as spin glass \cite{Queiroz06}, spin
liquid \cite{Zhou11}, and spin ice \cite{Bramwell09,Morris09,Fennell09}.
They attract a large amount of interests in solid state physics owing to
their intrinsic link to high temperature superconductivity. Stimulated by
these researches, we focus on a system of ultracold Rydberg atoms loaded
into a two-dimensional (2D) triangular optical lattice as shown in Fig. \ref%
{lattice}(a). The recent new progress in experiments could make it an ideal
system to investigate the effect of geometrical frustration in the near
future.

In our study, with the mean-field treatment, we investigate the steady state
phases of Rydberg atoms under strong dipole-dipole interactions (DDIs) \cite%
{Pillet08}. The dipole blockade effect results in a suppression of Rydberg
excitation for atoms in the nearest-neighbor (NN) lattice sites \cite%
{Viteau11}, which may give rise to a typical antiferromagnetic phase as
predicted in Ref. \cite{Tony11}. However, the triangular geometry adds
frustration effect into the scheme which results in very different quantum
phases from the square lattice case \cite{Jing12}. Except the uniform and
nonuniform phases that are stable when the laser detuning and DDIs are
dominated, respectively, bistable, oscillatory, and even chaotic phases are
predicted in this system. In addition, we also perform a full quantum
numerical simulation on a six-atom 2D model in a similar parameter region to
compare with the mean-field results.

\textit{Mean-field model}: We consider a 2D triangular lattice with exactly
one two-level atom per site. We assume that the lattice depth is deep enough
so the center-of-mass motion of the atoms can be safely neglected. An
off-resonant laser beam uniformly illuminates the atoms, driving the
transition between the atomic ground state $\left\vert g\right\rangle $ and
the Rydberg state $\left\vert r\right\rangle $. In the interaction picture,
the total Hamiltonian of the system reads $\mathcal{H}=\sum_{j}\mathcal{H}%
_{j}+\mathcal{V}\sum_{k\neq j}\left\vert r\right\rangle \left\langle
r\right\vert _{j}\otimes \left\vert r\right\rangle \left\langle r\right\vert
_{k}$, with $\mathcal{H}_{j}$ describing the atom at the $j$-th lattice site
and its coupling to the laser field,

\begin{equation}
\mathcal{H}_{j}=-\Delta \left\vert r\right\rangle \left\langle r\right\vert
_{j}+\frac{\Omega }{2}\left( \left\vert r\right\rangle \left\langle
g\right\vert _{j}+\left\vert g\right\rangle \left\langle r\right\vert
_{j}\right) ,  \label{Ham}
\end{equation}%
where $\Delta $ is the detuning between the laser frequency and atomic
transition frequency and $\Omega $ is the single-photon Rabi frequency. The
second term in total Hamiltonian $\mathcal{H}$ is for the DDIs between
Rydberg states of atoms. Since its strength decreases rapidly as the inverse
of $\left( a\left\vert j-k\right\vert \right) ^{6}$ with $a\left\vert
j-k\right\vert $ the distance between lattice sites $j$, $k$ and $a$ the
nearest distance, we truncate the summation to the NN sites \cite{Honing13}.
If we treat the two-level atom as a spin-$1/2$ particle, the total
Hamiltonian is formally equivalent to an Ising model with exchange energy $%
\mathcal{V}$ and transverse external field of strength $\Omega /2$ \cite%
{Elliott70}.

\begin{figure}[ptb]
\centering
\includegraphics[
height=1.8542in,
width=3.2993in
]%
{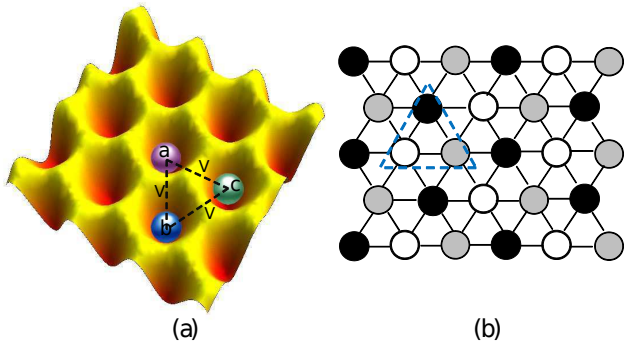}%
\caption{(Color online) (a) Schematic
diagram of the smallest triangular cell of three atoms (a, b and c) trapped
in a two-dimensional triangular lattice. The nearest-neighbor interatomic
interaction is denoted by $\mathcal{V}$. (b) This two-dimensional lattice is
composed of three interpenetrating sublattices marked in black, white and
gray colors, respectively. The dashed blue frame indicates the smallest
triangular unit.}
\label{lattice}%
\end{figure}

In order to obtain the quantum phases of the system, one should solve the
steady state solutions of the master equation:

\begin{equation}
\Dot{\rho}=-i\left[ \mathcal{H},\rho \right] +\mathcal{L}\left[ \rho \right]
,  \label{Master_eq}
\end{equation}%
where the spontaneous emission process from the Rydberg state with decay
rate $\gamma $ is described by the Lindblad operator $\mathcal{L}\left[ \rho %
\right] $,

\begin{equation}
\mathcal{L}\left[ \rho \right] =\gamma \left\vert g\right\rangle
\left\langle r\right\vert _{j}\rho \left\vert r\right\rangle \left\langle
g\right\vert _{j}-\frac{\gamma }{2}\left\{ \left\vert r\right\rangle
\left\langle r\right\vert _{j},\rho \right\} .  \label{Lind}
\end{equation}

For the 2D and large-number atomic system considered here, the mean-field
approximation (MFA) is a good treatment to simplify the master equation \cite%
{Diehl10}. Under the MFA, we can neglect the intersite quantum correlations
and factorize the density matrix each site. Then the master equation (\ref%
{Master_eq}) is reduced to the nonlinear coupling equations,

\begin{subequations}
\label{eqs_mas}
\begin{eqnarray}
\Dot{\rho}_{j,rr} &=&\Omega\ \text{Im}\left( \rho _{j,gr}\right) -\rho
_{j,rr}, \\
\Dot{\rho}_{j,gr} &=&-i\Delta _{j,eff}\rho _{j,gr}-\frac{1}{2}\rho _{j,gr}+i%
\frac{\Omega }{2}\left( 1-2\rho _{j,rr}\right) 
\end{eqnarray}%
where we have introduced the single-site density matrix elements $\rho
_{j,rr}$ and $\rho _{j,gr}$ to describe the $j$th atom's Rydberg state
population and atomic coherence, respectively. The effective detuning for
the $j$th atom, $\Delta _{j,eff}=\Delta -\mathcal{V}\sum_{k\neq j}\rho
_{k,rr}$ is renormalized by the excitation probabilities of its neighbors.
In arriving at Eqs. (\ref{eqs_mas}), the frequency is scaled with $\gamma $
and time with $1/\gamma $.

In order to include all the NN interactions and the effect of the
geometrical frustration, we factorize the triangular lattice into three
interpenetrating sublattices that are labeled by $j=1$ (black), $2$ (white)
and $3$ (gray) as shown in Fig.\ref{lattice}(b). In that way, only NN
interactions between Rydberg atoms from different sublattices are
considered, hence the steady state solutions $\rho _{j,rr}^{s}$ to Eqs. (\ref%
{eqs_mas}) are governed by the following equations:

\end{subequations}
\begin{equation}
\rho _{j,rr}^{s}=\frac{\Omega ^{2}}{4\left( \Delta _{j,eff}^{s}\right)
^{2}+1+2\Omega ^{2}},(j=1,2,3)  \label{coupled_eq}
\end{equation}%
with $\Delta _{j,eff}^{s}=\Delta -\mathcal{V}\sum_{k}\rho _{k,rr}^{s}$, $%
k=1,2,3$ and $k\neq j$. The superscript $s$ stands for the stationary
solution. In general, the quantum phases decided by $\rho _{j,rr}^{s}$ can
be classified into uniform and nonuniform phases. The uniform phase (UNI)
can also be called ferromagnetic phase, because it corresponds to a
spatially homogeneous Rydberg excitation probability, i.e. $\rho
_{1,rr}^{s}=\rho _{2,rr}^{s}=\rho _{3,rr}^{s}$. However, due to the
triangular geometry, the nonuniform phase has several different types, they
are:

a) Bi-antiferromagnetic phase (BAF): BAF phase corresponds to the case that
atoms in two of the three sublattices have the same Rydberg excitation
probability, e.g. $\rho _{1,rr}^{s}=\rho _{2,rr}^{s}\neq \rho _{3,rr}^{s}$.
This is similar to the "Y" state in classical triangular Heisenberg
antiferromagnet \cite{Sela11}. Physically, this originates from the
geometric frustration by which all the neighboring "spins" on the triangles
can not align antiparallel to each other at the same time. The ground state
is composed of two spins of one kind and one of another in each triangle.
When $\rho _{1,rr}^{s}=\rho _{2,rr}^{s}<\rho _{3,rr}^{s}$, we define it as
BAF1; when $\rho _{1,rr}^{s}=\rho _{2,rr}^{s}>\rho _{3,rr}^{s}$, we define
it as BAF2.

b) Tri-antiferromagnetic phase (TAF): TAF phase corresponds to exact
nonuniform steady-state solutions which requires $\rho _{1,rr}^{s}\neq \rho
_{2,rr}^{s}\neq \rho _{3,rr}^{s}$. This phase is a distorted version of the
"120-degree state" in Heisenberg antiferromagnet \cite{Sela11}.

\begin{figure}[ptb]%
\centering
\includegraphics[
height=1.4788in,
width=3.4618in
]%
{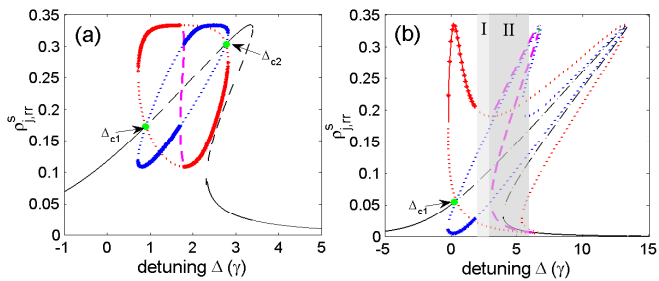}%
\caption{(Color online) Steady-state
solutions $\protect\rho _{j,rr}^{s}$ ($j=1,2,3$) versus detuning $\Delta $.
Stable and unstable uniform solutions are denoted by solid black and dashed
black lines. TAF solutions are denoted by dashed magenta lines. Stable and
unstable BAF solutions are denoted by thick solid lines with cross symbols
and thick dotted lines. The two same stationary solutions in BAF phase are
marked in blue color and the third one is marked in red color. (a) The weak
interaction case with $\mathcal{V}=5$; (b) The strong interaction case with $%
\mathcal{V}=20$. Light shadow regime I corresponds to the dynamical
oscillatory solutions, and dark shadow regime II to the chaotic behaviour.
Other parameter is $\Omega =1.0$. $\protect\gamma $ is the frequency unit.}%
\label{phase_diagram}%
\end{figure}

\textit{Phase diagram}: In Fig. \ref{phase_diagram}, we show the steady
state solutions $\rho _{j,rr}^{s}$ (j=1,2,3) and its stability with the
tunable one-photon detunings $\Delta $ at two different cases: (a) the weak
interaction case with $\mathcal{V}/\Omega =5$ and (b) the strong interaction
case with $\mathcal{V}/\Omega =20$. A similar and detailed demonstration for
the phases and their stabilities based on a cubic lattice with three-level
atoms has been presented in our recent work \cite{Jing12}. Here, we mainly
focus on the new findings for this triangular configuration.

In the weak interaction case, we find the system allows the existence of
rich stable phases that contains UNI, BAF1, BAF2 and TAF phases. By changing 
$\Delta $ from negative, there are two subcritical pitchfork bifurcations:
one is at $\Delta _{c1}=0.90$ in which the system undergoes a phase
tranition from UNI phase to BAF1 phase; the other is at $\Delta _{c2}=2.76$
in which the system transits back to UNI phase from BAF2 phase. The
subcritical pitchfork bifurcation means these phase transitions are
discontinuous, that is contrast to the case for square lattice \cite{Tony11}%
. In addition, the triangular geometry also enables a stable TAF phase at
the region $\Delta \in \left[ 1.70,1.79\right] $, which plays a role of
transitional phase between BAF1 and BAF2 phases. Moreover, bistability takes
place between two different UNI phases, UNI and BAF1 phases and UNI and BAF2
phases, which is a typical nonlinear effect due to the Rydberg interactions.

The impact of nonlinear Rydberg couplings combined with triangular geometry
becomes more prominent when turning to the strong interaction case.
Different from the weak interaction case, now the phase diagram [see Fig.\ref%
{phase_diagram}(b)] shows a clear three-peak character. From left to right,
the first peak located at $\Delta \approx 0$ corresponds to the single-atom
resonance which requires e.g. $\Delta _{3,eff}^{s}=0$, $\Delta
_{(1,2),eff}^{s}\neq 0$; the second at $\Delta =\mathcal{V}\Omega
^{2}/\left( 1+2\Omega ^{2}\right) $ corresponds to the double-atom resonance
which requires e.g. $\Delta _{(1,2),eff}=0$, $\Delta _{3,eff}\neq 0$; the
third at $\Delta =2\mathcal{V}\Omega ^{2}/\left( 1+2\Omega ^{2}\right) $
corresponds to the three-atom resonance which requires $\Delta
_{(1,2,3),eff}=0$. The stability of these stationary solutions are
numerically checked and unstable solutions are labeled by dashed and dotted
lines. Clearly, at single-atom resonance the stationary solutions are
stable, which means that the Rydberg blockade effect is very efficient \cite%
{Viteau12}. When $\Delta $ increases to satisfy two multi-atom resonance
conditions, the stationary solutions are both unstable, indicating
inefficient Rydberg antiblockade effect \cite{Ate07,Amthor10}. By solving
the resonances, we also note that $\max [\rho _{j,rr}^{s}]=\Omega
^{2}/\left( 2\Omega ^{2}+1\right) \rightarrow 1/2$ if $\Omega \rightarrow
+\infty $, which is consistent with the steady state solution of the optical
Bloch equations for an isolated two-level atom \cite{Meystre01}. In our
calculations, $\max [\rho _{j,rr}^{s}]=1/3$.

Besides, different from the weak interaction case, there are some parameter
regions without any stable phases. As $\Delta $ crosses the critical point $%
\Delta _{c1}$ and increases to the values in shadow region I, the Rydberg
state populations start to oscillate periodically in time [see Fig.\ref%
{Lyapunov exponent}(c)], until entering shadow regime II where the
population evolutions become irregular and complex [see Fig.\ref{Lyapunov
exponent}(d)]. We will show below that they indicate the emergence of chaos.
When $\Delta $ is even large blue-shifted, the system turns back to uniform
and low-excitation. To demonstrate the existence of BAF1 phase (not BAF2 or
TAF), we introduce a simple model with three atoms in which two of them
interact strongly, forming a typical two-atom Rydberg blockade model \cite%
{Comparat10}. As a third atom is brought closer towards this atomic pair, it
feels strong DDIs induced by the highly-excited atom in the pair and the
other ground state atom does not affect the third one. As a result, it is
only possible to obtain stable BAF1 phase in the strong interaction case.

It is worthwhile to state that the model we considered in this work is an
open system. The presence of driving and dissipation leads to remarkable
nonequilibrium phenomena in our model, such as periodic oscillation of
Rydberg excitation and bistability between different phases \cite{Glaetzle12}%
. Moreover, except for the support for various frustration phases, e.g. BAF
and TAF phases, the triangular geometry also plays a key role in generating
chaos in this open system, which is an important behavior in nonequilibrium
statistics and is absent in square lattice case.

\textit{Lyapunov exponent and Chaos: }The motions in the chaotic environment
could show very sensitive dependence to the initial conditions. This means
two trajectories starting very close to each other will rapidly diverge, and
can have totally different futures. To measure this sensitivity and to
verify that chaos is not a just long-period oscillation, we introduce a
parameter named "\textit{Lyapunov exponent}" \cite{Cencini09}. The presence
of at least one positive Lyapunov exponent value can be regarded as the
signature of chaos. For that, we introduce the definition of "\textit{%
maximal Lyapunov exponent"} (MLE):

\begin{equation}
\lambda _{\max }=\lim_{t\rightarrow \infty }\lim_{\delta Z_{0}\rightarrow 0}%
\frac{1}{t}\ln \left\vert \frac{\delta Z\left( t\right) }{\delta Z_{0}}%
\right\vert  \label{Lia_ex}
\end{equation}%
which describes two trajectories with initial small separation $\delta Z_{0}$
diverge as a function of $\delta Z\left( t\right) $ with time $t$.

\begin{figure}[ptb]%
\centering
\includegraphics[
height=2.4379in,
width=3.5206in
]%
{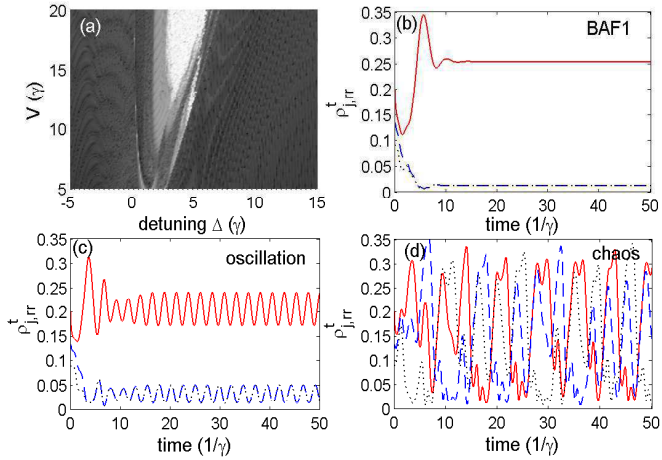}%
\caption{(Color online) (a) MLE values
in $(\Delta ,\mathcal{V})$ space with initial Rydberg populations $\protect%
\rho _{1,rr}^{t=0}=0.2$, $\protect\rho _{2,rr}^{t=0}=0.15$, $\protect\rho %
_{3,rr}^{t=0}=0.1$ and a small separation $\protect\delta \protect\rho %
_{1,rr}^{t=0}=10^{-6}$. The positive, near-zero and negative values are
respectively labeled by white, lightgray and darkgray colors, corresponding
to chaos, oscillatory phase and stable phase; (b)-(d) show typical Rydberg
population dynamics $\protect\rho _{1,rr}^{t}$ (solid), $\protect\rho %
_{2,rr}^{t}$ (dashed), $\protect\rho _{3,rr}^{t}$ (dotted) of BAF1 phase ($%
\Delta =1.0$), oscillatory phase ($\Delta =2.0$) and chaotic behavior ($%
\Delta =5.0$) with $\mathcal{V}=20$. Other parameters are $\Omega =1$, $%
\protect\gamma t=50$.}%
\label{Lyapunov exponent}%
\end{figure}

In Fig. \ref{Lyapunov exponent}(a) we plot the MLE values of $\rho _{1,rr}$
in the parameter space of $\left( \Delta ,\mathcal{V}\right) $ by
numerically solving Eqs. (\ref{eqs_mas}). Most of areas in darkgray
correspond to stable phases in which MLE is strictly negative. Such stable
phases include UNI, BAF1, BAF2 and TAF for weak interactions and UNI, BAF1
for strong interactions that is consistent with our former results. The
positive MLE values corresponding to the chaos only emerge in a small white
area, where the interaction strength $\mathcal{V}$ is larger than the
single-atom energy represented by the negative detuning $-\Delta $ and Rabi
frequency $\Omega $. Besides, there is also a lightgray area around the
chaotic area, whose MLE values are very close to zero. The existence of this
intermediate area is because when the system has a chaotic tendency, it
always first starts to oscillate quasi-periodically.

Figure \ref{Lyapunov exponent}(b)-(d) show the Rydberg-state population
evolutions $\rho _{j,rr}^{t}$ (j=1,2,3) for three sublattices in the BAF1
phase, oscillatory phase and chaotic regimes, respectively. Fig. \ref%
{Lyapunov exponent}(b) clearly shows that the system is stable at BAF1 phase
in which two atoms within a single triangular sublattice exhibit lower
excitation probabilities than the third one. This is due to the Rydberg
blockade effect. By increasing $\Delta $ to $2.0$ (in the lightgray regime
of (a)), the system enters into the oscillatory phase with the Rydberg
populations periodically oscillating surround its original fixed points [see
Fig.\ref{Lyapunov exponent}(c)]. For a larger detuning [Fig.\ref{Lyapunov
exponent}(d)], the system undergoes a transition to the chaotic motion where
the periodic oscillations become random and irregular. We note that there
are various ways for the dynamics of a classical system trends to be
chaotic, such as period-doubling oscillation \cite{Krefting09}, intermittent
chaos \cite{Shlizerman09}, and quasi-periodic oscillation \cite%
{Rand82,Zhang10} which is the approach for the present system.

\begin{figure}[ptb]%
\centering
\includegraphics[
height=2.7294in,
width=3.429in
]%
{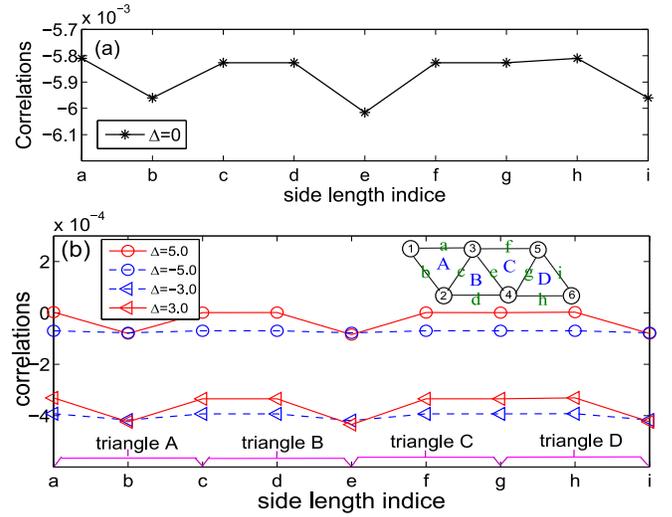}%
\caption{Quantum calculation solutions of
master equation for a six-atom lattice (inset of (b)) with periodic boundary
conditions. The correlation quantities $\protect\delta _{\protect\mu v}$ are
solved under different detunings. $a,b,...,i$ denote the side length indices
in four triangles $A$, $B$, $C$ and $D$. (a) $\Delta =0$ (solid stars); (b) $%
\Delta =-5$ (solid circles), $-3$ (dashed triangles), $3$ (solid triangles), 
$5$ (solid circles). Other paramters are the same as in Fig. \protect\ref%
{Lyapunov exponent}.}%
\label{Probability}%
\end{figure}

\textit{Full quantum simulation:} To complement the above analysis with MFA,
we present a full quantum simulation for a six-atom 2D lattice by solving
the original master equation (\ref{Master_eq}). The lattice includes four
triangular cells as shown in the inset of Fig. \ref{Probability}(b).
Following Ref. \cite{Tony11}, we introduce a quantity $\delta _{\mu
v}=\left\langle \sigma _{\mu }\sigma _{v}\right\rangle -\left\langle \sigma
_{\mu }\right\rangle \left\langle \sigma _{v}\right\rangle $ with $\sigma
_{\mu }=\left\vert r\right\rangle \left\langle r\right\vert _{\mu }$ and $%
\mu \left( v\right) $ the atomic indices in each cell. This quantity
characterizes the quantum correlations between any two atoms in each
triangular cell and intimately related to the properties of this system.

In Figure \ref{Probability} we compare $\delta _{\mu v}$ for different
values of detuning $\Delta $. For $\Delta <0$ (dashed blue line), we observe
the difference between each $\delta _{\mu v}$s is very small, which indicate
the uniform phase obtained in MFA treatment. The small negative value of all 
$\delta _{\mu v}$s indicates the anti-correlation between NN atoms due to
the fact that the probability of two atoms being excited individually is
larger than that they are being excited simultaneously, i.e. Rydberg
blockade effect. For $\Delta =0$ [Fig. \ref{Probability}(a)], we see that in
each triangle [A, B, C, D], two of the three $\delta _{\mu v}$s are always
very close and different from the third one, that confirms the prediction of
BAF1 phase with MFA at the same parameters. As $\Delta $ increases, we find $%
\delta _{\mu v}$ tends to zero (around $\Delta =5$) from negative but the
difference between $\delta _{\mu v}$s still remains. That shows a tendency
of transition from BAF1 to BAF2 phase. With the further increase of $\Delta $%
, $\delta _{\mu v}$ become positive, corresponding to BAF2 phase due to the
anti-blockade effect (not show in the figure).

In our full quantum calculation, the oscillatory and chaotic phase are not
observed. That is because these nonequilibrium phases originates from the
nonlinear coupling between the density matrix elements of the NN atoms in
MFA equations (\ref{eqs_mas}) [see the definition of $\Delta _{j,eff}$] \cite%
{Strogatz94}. However, the present quantum model consists of only six atoms,
which is obviously far from the effective regime of MFA, so that it can not
display the quantum counterparts of these phases. Actually, in order to
investigate the corresponding quantum behaviors of a chaotic system, its
quasi-probability distribution in phase space should be calculated \cite%
{Gardiner97}. This problem will be addressed in our future works.

\textit{Conclusions}: We have investigated quantum phases of strongly
interacting Rydberg atoms in a 2D triangular lattice system via MFA. The
effect of laser pumping and spontaneous decay are considered. Except the
uniform phase, we find that the geometric frustration results in rich
phases. By tuning the pump detuning, the system shows BAF phase, TAF phase
and even bistable states between these phases. Particularly, in the strong
interaction case, the system has no stable phase in some parameter region,
in which its dynamics is characterized by chaotic oscillations. To confirm
the existence of chaos, we calculate the maximal Lyapunov exponent and find
that it is strictly positive in the chaotic regime. Moreover, we present a
full quantum calculation with six atoms to simulate the Rydberg excitation
probability and find that it is consisent with the mean-field results.

Our results show that the Rydberg atoms in a triangular optical lattice is a
good candidate system to simulate the quantum spin frustration and explore
its effect in the nonequilibrium statistics and quantum chaos. In the future
works, we will extend our discussion to Rydberg atoms with long-range and
anisotropic DDIs, which play the important roles in the generation of spin
ice, spin glass and other unique quantum phases.

This work was supported by the National Basic Research Program of China (973
Program) under Grant No. 2011CB921604, the National Natural Science
Foundation of China under Grants Nos. 11104076, 11004057, 11234003, the
Specialized Research Fund for the Doctoral Program of Higher Education No.
20110076120004, the "Chen Guang" project from Shanghai Municipal Education
Commission and Shanghai Education Development Foundation under Grant No.
10CG24, Shanghai Rising-Star Program under Grant No. 12A1401000 and the
Fundamental Research Funds for the Central Universities.

\end{document}